\documentclass[aps,reprint,article,prx]{revtex4-1}
\usepackage{amsmath}
\usepackage{dcolumn}
\usepackage{graphicx}
\usepackage{float}

\begin{document}

\title{Direct visualization of sign reversal $s^\pm$ superconducting gaps in FeTe$_{0.55}$Se$_{0.45}$}

\author{Mingyang Chen, Qingkun Tang, Xiaoyu Chen, Qiangqiang Gu, Huan Yang,$^*$ Zengyi Du, Xiyu Zhu, Enyu Wang, Qiang-Hua Wang,$^\dag$ and Hai-Hu Wen$^\ddag$}

\affiliation{National Laboratory of Solid State Microstructures and Department of Physics, Collaborative Innovation Center of Advanced Microstructures, Nanjing University, Nanjing 210093, China}

\begin{abstract}
In many unconventional superconductors, the pairing of electrons is driven by the repulsive interaction, which leads to the sign reversal of superconducting gaps along the Fermi surfaces (FS) or between them. However, to measure this sign change is not easy and straightforward. It is known that, in superconductors with sign reversal gaps, non-magnetic impurities can break Cooper pairs leading to the quasiparticle density of states in the superconducting state. The standing waves of these quasiparticles will interfere each other leading to the quasiparticle interference (QPI) pattern which carries the phase message reflecting also the superconducting gap structure. Based on the recently proposed defect-bound-state QPI technique, we explore the applicability of this technique to a typical iron based superconductor FeTe$_{0.55}$Se$_{0.45}$ with roughly equivalent gap values on the electron and hole pockets connected by the wave vector $\mathbf{q}_2=(0,\pi)$. It is found that, on the negative energy side, with the energy slightly below the gap value, the phase reference quantity $|g(\mathbf{q},-E)|\cos(\theta_{\mathbf{q},+E}-\theta_{\mathbf{q},-E})$ becomes negative and the amplitude is strongly enhanced with the scattering vector $\mathbf{q}_2$, but that corresponding to the scattering between the electron-electron pockets, namely $\mathbf{q}_3=(\pi,\pi)$, keeps all positive. This is well consistent with the theoretical expectation of the $s^\pm$ pairing gap and thus serves as a direct visualization of the sign reversal gaps. This experimental observation is also supported by the theoretical calculations with the Fermi surface structure and $s^\pm$ pairing gap.
\begin{description}
\item[Subject Areas] Condensed Matter Physics, Strongly Correlated Materials, Superconductivity
\end{description}
\end{abstract}

\maketitle
\section{Introduction}
Superconductivity is originated from the condensation of Cooper pairs based on the Bardeen-Cooper-Schrieffer (BCS) theory. In conventional superconductors, as predicted by the BCS theory, a Cooper pair is induced by the attraction between two electrons through exchanging phonons. However, the pairing interaction can also be repulsive in some unconventional superconductors, such as high-$T_c$ cuprates, which leads to the sign reversal of the superconducting (SC) gap(s). Practically, to determine the sign change of the superconducting gap is not an easy task. For example, in cuprates, to confirm the sign-reversal $d$-wave gap, several sophisticated phase-sensitive experiments on different kinds of Josephson junctions were designed and conducted \cite{TsueiKirtleyReview}. The iron-based superconductors are the second family of unconventional high-$T_c$ superconductors, and the widely accepted picture is $s^\pm$ pairing mediated by the antiferromagnetic spin fluctuations \cite{MazinS+-,KurokiS+-}. However, the phase-sensitive technique based on Josephson junctions seems not working in iron-based superconductors because of the roughly isotropic superconducting gap on each Fermi pocket \cite{HirshfeldReview}. A theoretical design for phase-sensitive experiment was proposed based on a sandwich-like junction containing both hole- and electron-doped films along $c$-axis for iron-based superconductors \cite{sandwich}, but the experimental results, as far as we know, are still lacking. Nevertheless, there are still several experiments supporting the $s^\pm$ pairing in iron-based superconductors with both electron and hole pockets. One of the major experiments is the observation of a resonance peak of the magnetic excitation at ($\pi$,$\pi$) in the SC state by the inelastic neutron scattering experiment \cite{Neutron}. Furthermore, there are some indications for the $s^\pm$ pairing from the scanning tunnelling microscopy (STM) measurements. These include the observation of the momentum resolved intensity change of Fourier transformed (FT-) quasiparticle interference (QPI) by using a magnetic field in Fe(Se,Te) \cite{HanaguriQPI}, and the in-gap bound states induced by non-magnetic Cu impurities in Na(Fe$_{0.96}$Co$_{0.03}$Cu$_{0.01}$)As \cite{YangHNC}. Moreover, the STM experiments in different FeAs-based systems reveal the existence of bosonic modes with energies identical to the neutron spin resonance energies, this can also be naturally regarded as the consequence of the sign-reversed gap \cite{WangZYNP}.

The QPI method, which is designed for measuring the spatial evolution of the differential conductance $g(\mathbf{r},E)$, is a very useful tool in STM measurement and can provide essential message of electronic properties. The quasiparticles in a metal can be scattered by defects, impurities or boundaries on the surface. The resultant standing waves will interfere each other and show some special patterns in the QPI images $g(\mathbf{r},E)$ which reflect complex information related to the Fermi surface. When the Fourier transform is operated on QPI image $g(\mathbf{r},E)$, the resultant Fourier transformed (FT-) QPI signal is a complex function which can be expressed as $g(\mathbf{q},E)=|g(\mathbf{q},E)| \exp(i\theta_{\mathbf{q},E})$ with $\theta_{\mathbf{q},E}$ the phase. The intensity of the FT-QPI patterns $|g(\mathbf{q},E)|$ is proportional to the joint density of states (DOS) between the scattering momenta ($\mathbf{k}_1$ and $\mathbf{k}_2$) on the Fermi surfaces with the momentum difference of $\mathbf{q}=\mathbf{k}_2-\mathbf{k}_1$ \cite{HoffmanReivew}. The QPI technique is also used to detect the Bogoliubov quasiparticles in a superconductor, and it will provide rich information of the superconducting gap values \cite{HanaguriQPI,AllanScience,HessPRL,ChiPRB,DuNC,ChenMYSA}. It should be noted that FT-QPI data are all complex values which contain the phase information. The phase $\theta(\mathbf{q},E)$ of FT-QPI has close relationship with the detailed structure of the superconducting gap in superconducting state. Therefore the phase-referenced (PR-)QPI method was proposed theoretically by Hirschfeld, Altenfeld, Eremin, and Mazin (HAEM) \cite{PR-QPI}. This method was initially designed for detecting the sign of two SC gaps ($\Delta_1$ and $\Delta_2$) for the case of single non-magnetic impurity. The key factor of this method is the antisymmetrized FT-QPI signal
\begin{equation}
\delta \rho^-(E) =\sum_{q\in A}\delta \rho^-(q,E)=\sum_{q\in A}[g(\mathbf{q},+E)-g(\mathbf{q},-E)].\label{eqHAEM}
\end{equation}
The summation is taken in the area $A$ of all $q$ scattering vectors. The PR-term $\delta \rho^-(E)$ will manifestly enhance at the energies between $\Delta_1$ and $\Delta_2$ for sign-reversal gaps, and $\delta \rho^-(E)$ will be very small with sign alternating at the energies between $\Delta_1$ and $\Delta_2$ for gaps with the same sign. This method was successfully used to determine the orbital selected sign-reversal pairing gaps in FeSe \cite{DavisFeSe} with both electron and hole pockets, as well as in (Li$_{1-x}$Fe$_{x}$)OHFe$_{1-y}$Zn$_{y}$Se \cite{DuZYNP} with only electron pockets. This prediction, according to our understanding, may also work for a particular momentum $q$, in this case, eq.\ref{eqHAEM} changes into
\begin{equation}
\delta \rho^-(q,E)=g(\mathbf{q},+E)-g(\mathbf{q},-E).\label{eqHAEM2}
\end{equation}
Recently, another defect-bound-state (DBS-) QPI method was proposed \cite{ChiSTheory,ChiSExp}, which is effective near the impurity-induced bound-state energies within the SC gaps. According to the descriptions of this method, the PR-QPI signals at positive and negative energies are expressed as
\begin{eqnarray}
 g_{pr}(\mathbf{q},+E) &=& |g(\mathbf{q},+E)|\times\mathrm{Re}[e^{i(\theta_{\mathbf{q},+E}-\theta_{\mathbf{q},+E})}]\nonumber\\
 &=& |g(\mathbf{q},+E)|,\label{eq1}\\
g_{pr}(\mathbf{q},-E) &=& |g(\mathbf{q},-E)|\times\mathrm{Re}[e^{i(\theta_{\mathbf{q},-E}-\theta_{\mathbf{q},+E})}]\nonumber\\
&=& |g(\mathbf{q},-E)|\cos(\theta_{\mathbf{q},-E}-\theta_{\mathbf{q},+E}).\label{eq2}
\end{eqnarray}
Here Re is the operator to compute the real part of a complex number, $g_{pr}(\mathbf{q},+E)$ is always positive according to the definition, and $\theta_{\mathbf{q},-E}-\theta_{\mathbf{q},+E}$ is the phase difference of the FT-QPI signals at positive and negative energies. In their theoretical model \cite{ChiSTheory}, $g_{pr}(\mathbf{q},-E)$ should be negative near the bound state energy if the impurity is non-magnetic in a superconductor with sign-reversed gap, while the value is positive for the situation of a magnetic impurity on a sign-preserved gap. This DBS-QPI method was used to probe the sign change of the SC order parameter in LiFeAs \cite{ChiSExp}, and it has also been successfully used to confirm the sign-change of gaps in (Li$_{1-x}$Fe$_{x}$)OHFe$_{1-y}$Zn$_{y}$Se \cite{GuQQPRB}. We also applied the DBS-QPI method to the cuprate system Bi$_2$Sr$_2$CaCu$_2$O$_{8+\delta}$ (Bi2212) which does not contain any strong impurity bound state peak, the results clearly show the sign-change of the $d$-wave gap in the system.\cite{QQGuBi2212}.

In this work, we measured and analysed the FT-QPI patterns in the superconductor FeTe$_{0.55}$Se$_{0.45}$ with shallow and multiple bands. It is found that the DBS-QPI method can be successfully applied to determine and visualize the sign reversal gaps in this superconductor although it has no clear strong bound state peak. We also argue that the two PR-QPI methods mentioned above can reach the same conclusion about the gap sign problem in a qualitative way. Our results indicate that the DBS-QPI method can also work successfully even for the system which does not have bound state peaks, this may serve as an extension of this method to many other systems.

\section{Results}

\subsection{Tunneling spectra and QPI measurements on FeTe$_{0.55}$Se$_{0.45}$}

\begin{figure}
\includegraphics[width=9cm]{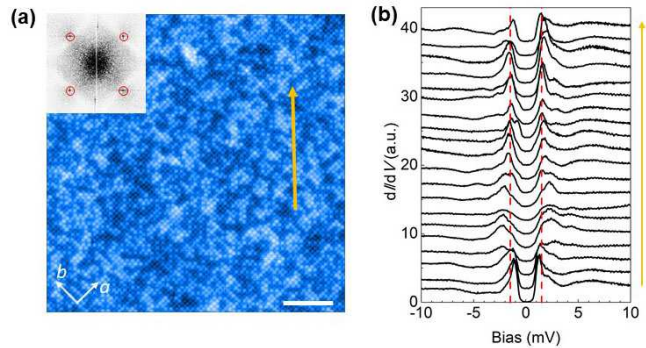}
\caption{Topographic image and tunneling spectra on FeTe$_{0.55}$Se$_{0.45}$. (a) Atomically-resolved topographic image with the square lattice on the cleaved surface measured with a bias voltage of $V_{bias} = 20$ mV and tunneling current of $I_{set} = 200$ pA. The inset is the Fourier transform result of the topographic image, the four spots marked by red circles are the Bragg peaks of the Se-Se crystalline lattice on the surface. Scale bar, 5 nm. (b) Spatially resolved tunneling spectra measured along the arrowed line at 0.4 K. The dashed lines denote the bias voltage values of $\pm 1.5$ mV, respectively. The spectrum features are inhomogeneous near the coherence peaks.
} \label{fig1}
\end{figure}

Figure~\ref{fig1}(a) shows a typical topography measured on FeTe$_{0.55}$Se$_{0.45}$. The atoms resolved on the top surface form a square lattice, which can also be reflected by the fourfold symmetric Bragg peaks marked by red circles in the FT image of the topography shown in the inset of Fig.~\ref{fig1}(a). The brighter atoms are the Te atoms with larger atomic sizes, while the darker ones are the Se atoms according to previous reports \cite{HanaguriQPI,ACSNano,YinJXNP,DuZYActa}. We have not observed any interstitial Fe atoms on the cleaved surfaces, which may manifest that most of the interstitial Fe impurities have been removed by the annealing treatment (see Appendix A). Figure~\ref{fig1}(b) shows spatial evolution of the tunneling spectra measured along the arrowed line in Fig.~\ref{fig1}(a). The spectra show a fully gapped feature with pairs of coherence peaks with peak energies at about $\pm 1.5$ mV. From our previous report \cite{ChenMYCdGM}, the tunneling spectra are very inhomogeneous in FeTe$_{0.55}$Se$_{0.45}$, and the energies of the coherence peaks range from about $\pm 1.1$ mV to $\pm 2.1$ mV. The spectrum shown here also varies slightly when traveling along the arrowed line, reflecting the spatially inhomogeneous electronic properties.

\begin{figure*}
\includegraphics[width=14cm]{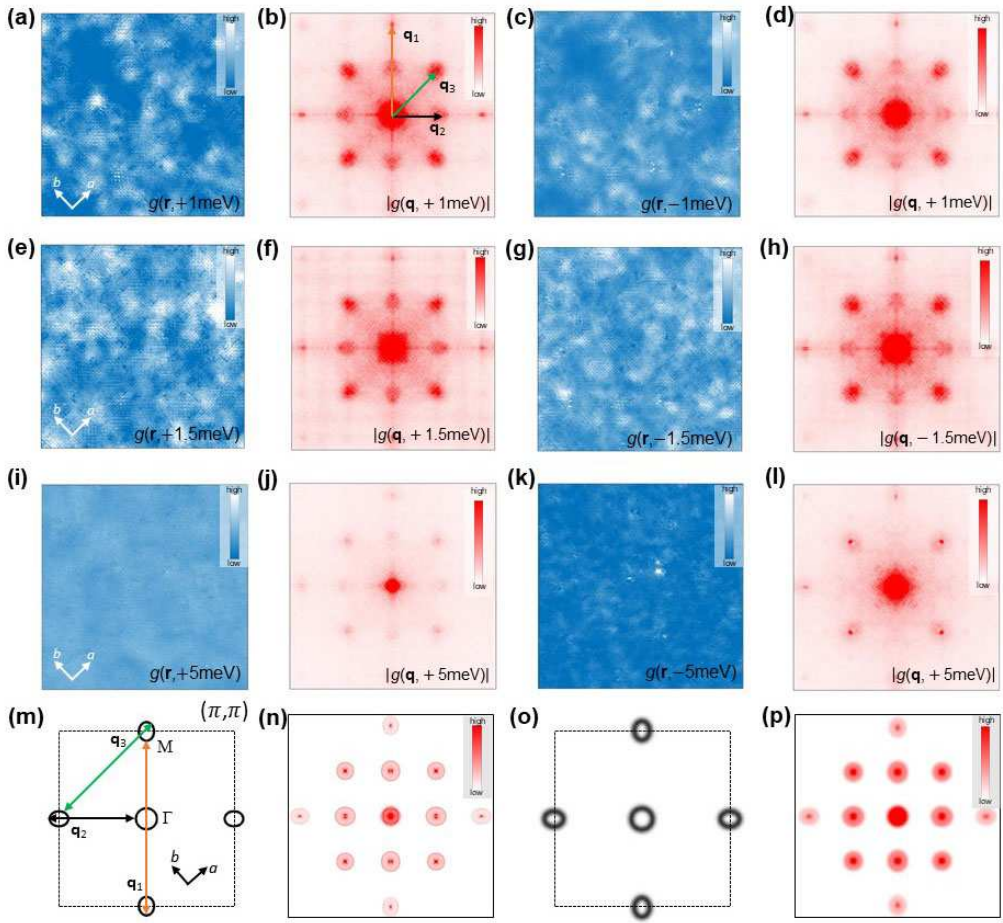}
\caption{(a,c) QPI images $g(\mathbf{r},E)$ in real space measured at 0.4 K and $\pm 1$ meV ($V_{set} = 10$ mV, $I_{set} = 200$ pA). (b,d) The FT-QPI patterns $|g(\mathbf{q},E)|$ derived from Fourier transform to the QPI images in (a,b), respectively. The inter-Fermi-pocket scattering vectors of QPI patterns are marked by $\mathbf{q}_1$, $\mathbf{q}_2$ and $\mathbf{q}_3$ in (b). (e-l) QPI and FT-QPI images measured at $\pm 1.5$ and $\pm 5$ meV ($V_{set} = 10$ mV, $I_{set} = 200$ pA). All the FT-QPI patterns have been fourfold symmetrized to enhance the signal-to-noise ratio. (m,o) Schematic plots of the elliptic-shaped electron pockets around M point and circular-shaped hole pocket around $\Gamma$ point in unfold Brillouin zone. Here each Fermi pocket has a certain width which is constructed by assuming a Gaussian form of DOS crossing the line, this serves as a purpose for simulation (see APPENDIX B). (n,p) Simulated FT-QPI patterns by applying self-correlations to (m) and (o), respectively. The scattering vectors are also plotted in (m). $\mathbf{q}_1$ and $\mathbf{q}_3$ are inter-electron-pocket scatterings, while $\mathbf{q}_2$ is the scattering betweeen electron- and hole-pockets.} \label{fig2}
\end{figure*}

We then perform the QPI measurements in the area shown in Fig.~\ref{fig1}(a) from $-6$ mV to $+6$ mV in order to obtain the information of gap symmetry. The results are shown in Fig.2. Figures~\ref{fig2}(a,c) and \ref{fig2}(e,g) show the typical QPI images measured at $\pm 1.0$ mV and $\pm 1.5$ mV, respectively. The standing waves can be clearly observed along Se-Se or Fe-Fe directions in these images. It is known that the modulations of these standing waves with different periodicity and along different directions will give rise to distinct features in $\mathbf{q}$-space. The corresponding FT-QPI patterns are shown in Figs.~\ref{fig2}(b,d) and \ref{fig2}(f,h). The two scattering spots near $\mathbf{q}_2$ and $\mathbf{q}_3$ have much stronger intensities compared to the one near $\mathbf{q}_1$ because of the shorter scattering wave-vectors. It should be noted that the spots with the center at $\mathbf{q}_3$ coincide with Se-Se Bragg peaks, which will enhance the intensities near the center of the $\mathbf{q}_3$ spot. The scattering intensities of all the patterns are weakened when the energy exceeds the superconducting gap [e.g. see Fig.~\ref{fig2}(i-l) measured at $\pm5$ meV], and only Se-Se Bragg peaks are left then. Therefore, the scattering is strengthened by the Bogoliubov quasiparticles near and below the superconducting gap energy.

In order to have a basic understanding on the FT-QPI patterns, in Fig.~\ref{fig2}(m) we plot the schematic Fermi surface structure of FeTe$_{0.55}$Se$_{0.45}$ based on the angle-resolved photoemission spectroscopy (ARPES) data \cite{FeSeTeARPES1}. Here we only plot one hole pocket around $\Gamma$ point for simplification. There are three inter-pocket scattering channels from the topology of Fermi surfaces, i.e., one associated with the scattering between the hole- and electron-pockets (marked by $\mathbf{q}_2$) and the other two associated with the scattering between electron pockets (marked by $\mathbf{q}_1$ and $\mathbf{q}_3$). For simulating the FT-QPI, we assume a Gaussian form DOS distribution crossing the line of the Fermi pockets, and the details can be found in APPENDIX B. By doing the self-correlation to Fig.~\ref{fig2}(m), we can obtain the simulated FT-QPI pattern as shown in Fig.~\ref{fig2}(n). Having a glance at the experimental data, however, one finds that all the scattering spots around the relevant wave vectors ($\mathbf{q}_1$, $\mathbf{q}_2$, and $\mathbf{q}_3$) in the experiment data show up as blurred pads and do not exhibit the clear outer contour, which is different from the simulated pattern in Fig.~\ref{fig2}(n). This observation is consistent with previous report \cite{HanaguriQPI}. The outer contour of FT-QPI was however observed in LiFeAs \cite{AllanScience} and (Li$_{1-x}$Fe$_{x}$)OHFeSe \cite{DuNC}. We attribute the missing of the outer contour of the FT-QPI spots to shallow bands and small Fermi pockets in FeTe$_{0.55}$Se$_{0.45}$. When we refer the band structure observed by ARPES measurements in FeTe$_{0.55}$Se$_{0.45}$ \cite{FeSeTeARPES2}, the Fermi energies are only several meV for both hole and electron pockets, which is confirmed by the analysis on the discrete vortex-bound-state results from STM measurements \cite{ChenMYCdGM}. Thus the Fermi pockets in FeTe$_{0.55}$Se$_{0.45}$ are much smaller than the Fermi pockets in LiFeAs \cite{AllanScience} and (Li$_{1-x}$Fe$_{x}$)OHFeSe \cite{DuNC}. In addition, for a shallow electron band, the slope of $dk/dE$ is very large, within the fixed energy window around Fermi energy, the Fermi surface is strongly smeared with a shape of thick belt. This argument is verified by ARPES data, and the Fermi pockets seem to be very blurred and thick in FeTe$_{1-x}$Se$_{x}$ \cite{FeSeTeARPES1,FeSeTeARPES2}. Based on this consideration, we did a further simulation with a thicker Fermi pocket by varying the Gaussian distribution function, the related results are shown in Fig.~\ref{fig2}(o,p). One can see that the edges of the scattering patterns become blurred if the line width of the Fermi surface contour is large enough. To further check this idea, we did a simulation with the same thickness of the Fermi pocket line width, but two times larger size of Fermi pockets. The results are presented in Fig.~\ref{fig6} of APPENDIX B. One can see that the outer contour of the scattering spots becomes very clear. So we conclude that the blurred boundaries of the scattering spots in FeTe$_{0.55}$Se$_{0.45}$ may be originated from the effect of shallow band and small Fermi pockets in this material.

\subsection{DBS-QPI technique applied on FeTe$_{0.55}$Se$_{0.45}$}

\begin{figure}
\includegraphics[width=8cm]{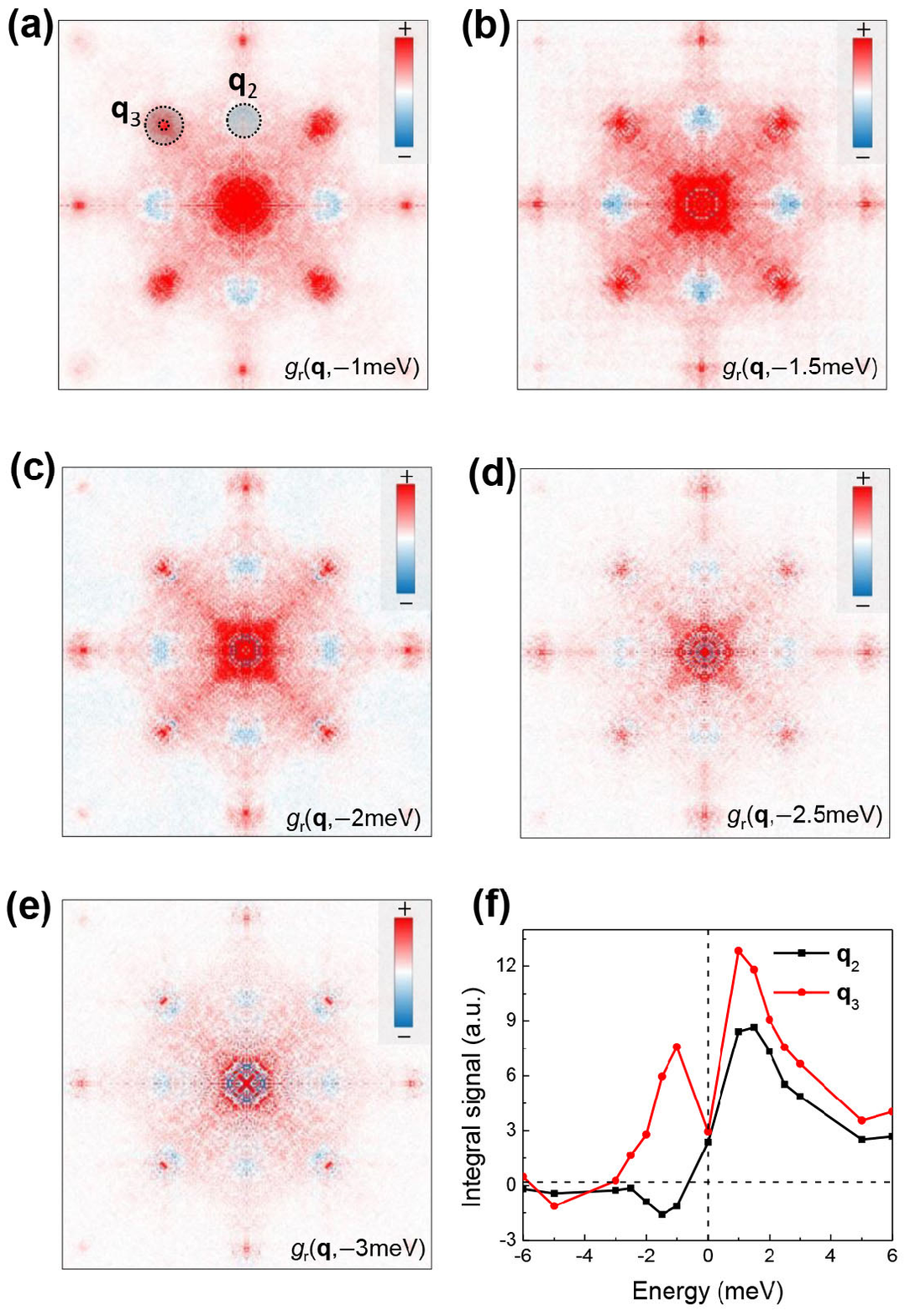}
\caption{PR-QPI patterns and the integral signal by DBS-QPI technique. (a-e) PR-QPI patterns $g_{pr}(\mathbf{q},-E)$ at energies $-E$ from $-1.0$ to $-3.0$ meV. The images are fourfold symmetrized to enhance the signal-to-noise ratio. The shaded regions near $\mathbf{q}_2$ and $\mathbf{q}_3$ are used as integral areas, and the central parts near $\mathbf{q}_3$ are excluded because it contains the intensity of Bragg peaks. (f) Energy dependence of the calculated integral signal of $g_{pr}(\mathbf{q},- E)$ within the shaded regions.
} \label{fig3}
\end{figure}

The FT-QPI patterns shown in Fig.~\ref{fig2} contain only the information of the intensities of all the scatting spots, which cannot tell any information of superconducting gap signs. For a superconductor with sign change between different Fermi surfaces, there will be additional phase shift induced by the non-magnetic impurity near the bound state peak energies \cite{ChiSTheory}. However, we can not see any clear in-gap impurity-induced bound-state peaks from the spectra in Fig.~\ref{fig1}(b), and the only feature is the spatial inhomogeneity of the coherence peaks. We then apply the DBS-QPI technique to the QPI data at several energies in order to obtain additional phase information of the superconducting gap. For that purpose, we first obtain the FT-QPI results for each energy $E$. Since for each pair of $\mathbf{q}$ and $E$, the output of FT-QPI is a complex quantity which contains both amplitude $|g(\mathbf{q},E)|$ and phase $\theta_{\mathbf{q},E}$. The PR-QPI $g_{pr}(\mathbf{q},\pm E)$ patterns can be obtained by applying Eqs.~\ref{eq1} and \ref{eq2} to the FT-QPI results. According to Eq.~\ref{eq1}, the $g_{pr}(\mathbf{q},+ E)$ patterns are the same as FT-QPI $|g(\mathbf{q},+ E)|$, thus we only show the $g_{pr}(\mathbf{q},- E)$ patterns in Fig.~\ref{fig3}(a-e). When energy is taken from 1 to 2 meV, the scattering spots from Bogoliubov quasiparticles are very clear with scattering vectors near $\mathbf{q}_1$, $\mathbf{q}_2$ and $\mathbf{q}_3$, but with different behaviors for positive and negative energies. It is obvious that the $g_{pr}(\mathbf{q},- E)$ values for scattering spots near $\mathbf{q}_2$ are negative, while the ones around $\mathbf{q}_3$ or $\mathbf{q}_1$ are positive. The PR signal $g_{pr}(\mathbf{q},- E)$ for scattering between electron- and hole-pockets exhibit clear opposite sign, which is in sharp contrast with the positive values for scattering between electron pockets. This strongly indicates that the gap changes sign between hole and electron pockets, being fully consistent with the theoretically proposed $s^\pm$ pairing picture. We also integrate the $g_{pr}(\mathbf{q},- E)$ signal in the shaded areas near $\mathbf{q}_2$ and $\mathbf{q}_3$ as shown in Fig.~\ref{fig3}(a), and the energy evolution of the integral signal is shown in Fig.~\ref{fig3}(f). There are positive and negative peaks locating at about $\pm 1.5$ meV which is close to the gap value of the sample as inferred from Fig.~\ref{fig1}(b) and our previous statistics \cite{ChenMYCdGM}. When the energy is much larger than the superconducting gap, for example as shown in Fig. ~\ref{fig3}(e) for 3 meV, the signal from the inter-pocket scatting become very weak, leaving mainly positive signal of Se-Se Bragg peaks in the PR-QPI patterns. The above results and analysis indicate that we can successfully determine and visualize the sign-reversal superconducting order parameters in FeTe$_{0.55}$Se$_{0.45}$ even without any obvious impurity-induced bound state peaks as requested by the originally proposed DBS-QPI technique. This may serve as an extension of this useful technique.

When the superconducting gaps change signs, the DBS-QPI technique \cite{ChiSTheory} expects a negative peak of $g_{pr}(\mathbf{q},- E)$ accompanying with a positive peak of $g_{pr}(\mathbf{q},+E)$ with peak energies near the non-magnetic impurity bound-state energies $\pm E_B$. Therefore, it is a surprise that there is a negative peak around the gap energy for the $g_{pr}(\mathbf{q},- E)$ data in FeTe$_{0.55}$Se$_{0.45}$. From our previous work in Bi2212  \cite{QQGuBi2212}, we found that, if there are widely distributed non-magnetic impurities with very weak scattering potential, the DBS-QPI technique can also be used to determine the sign reversal of SC gap. The situation in FeTe$_{0.55}$Se$_{0.45}$ may be similar to the case in Bi2212, the bound state peaks induced by the non-magnetic impurities may mix with the coherence peaks. For the sample we measured here, the interstitial Fe impurities have been removed by the annealing treatment, and we do not observe any Fe impurities from the topography measured by STM. Thus we believe that most of the impurities which induce the QPI scatterings are non-magnetic in nature. Theoretically, the bound-state peaks induced by non-magnetic impurities locate near zero-bias only when the impurity scattering potential $V_s$ has suitable values, e.g., $~$1 eV in iron pnictides \cite{OgataImpurity}. For weaker scattering potentials, the bound states may appear near the gap edges and mix with the coherence peaks. In the situation of FeTe$_{0.55}$Se$_{0.45}$, it is obvious that the coherence peaks are very inhomogeneous\cite{ChenMYCdGM}, which may be originated from the mixture of the coherence peak and some intensities of bound states. Next we do the theoretical calculations to prove this idea.

\subsection{Theoretical calculation results}

\begin{figure*}
\includegraphics[width=16cm]{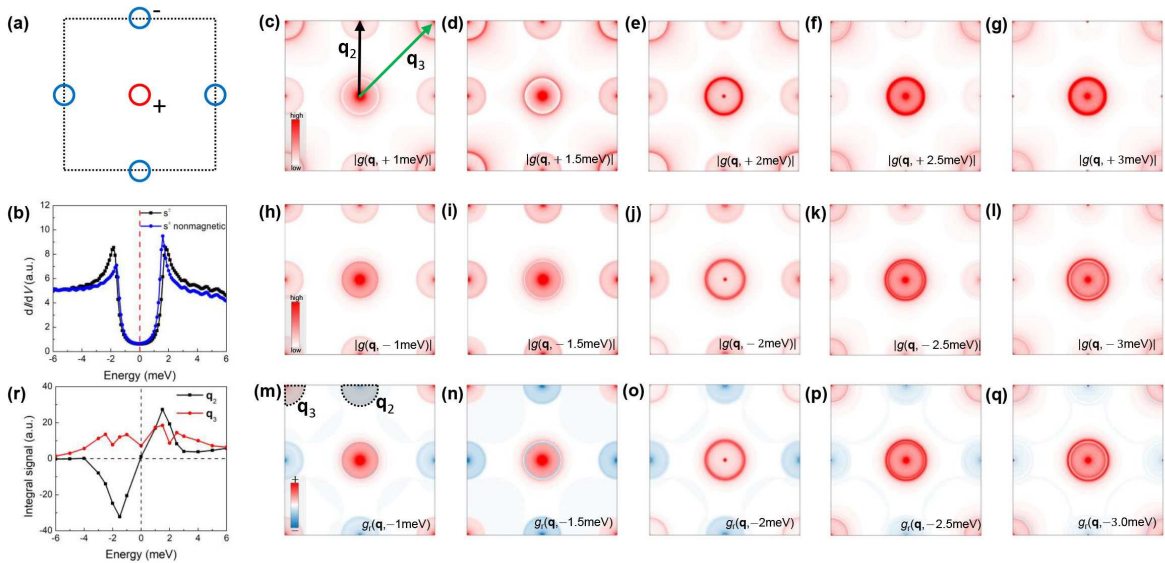}
\caption{Theoretical results of PR-QPI and the analysis by DBS-QPI method applied on a single non-magnetic impurity. (a) Schematic plot of the calculated Fermi surface with two-orbital model. The dashed black square denotes the unfold Brillouin zone with one Fe unit cell. (b) The calculated tunneling spectra with and without non-magnetic impurity, and the scalar potential $V_s=-24$ meV for the non-magnetic impurity. (c-l) The calculated FT-QPI patterns at different energies in a superconductor with $s^\pm$ pairing. (m-q) The simulated PR-QPI patterns by the DBS-QPI technique. (r) The integral signal of $g_{pr}(\mathbf{q},- E)$ as a function of energy. The integration areas are in the shaded regions shown in (m) for the patterns near $\mathbf{q}_2$ and $\mathbf{q}_3$.
} \label{fig4}
\end{figure*}

\begin{figure}
\includegraphics[width=8cm]{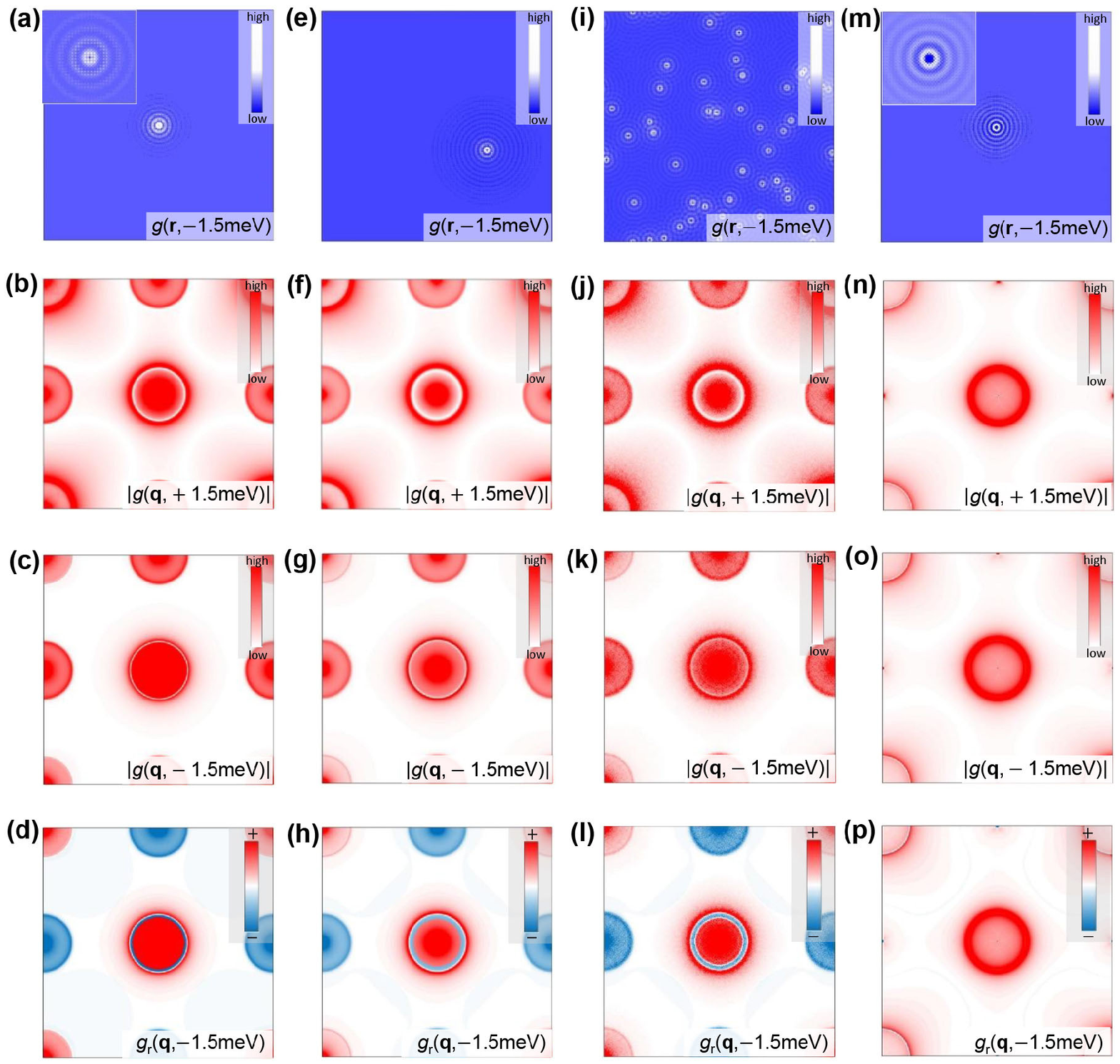}
\caption{The calculated PR-QPI patterns for different situations. (a) The calculated local DOS (LDOS) image (corresponding to the QPI image from experiment), (b,c) FT-QPI patterns, and (d) PR-QPI pattern when a single non-magnetic impurity locates at the center of the image in an $s^\pm$ superconductor. The parameters for the calculation are the same as the ones used for the calculating results in Fig.~\ref{fig4}. The inset in (a) shows an enlarged view of the calculated LDOS image near the impurity. (e-h) The calculated results by moving the impurity away from the center. (i-l) The simulated results for 50 randomly distributed impurities which have the same scattering potential as the one used in (a). The display area is also the same as that in (a), and the QPI image in (i) is from the direct superposition of QPI images of all the 50 impurities. (m-p) The simulated results for $s^{++}$ pairing when a single non-magnetic impurity locates at the center of the image. The parameters for the calculation are in APPENDIX C. The inset of (m) shows the enlarged view of LDOS image with the impurity, and the shape is a little different from the image shown in the inset of (a) although the scalar potential are both $V_s=-24$ meV.
} \label{fig5}
\end{figure}

We use a simple two-orbital model to do the numerical simulation, and the details of the theoretical calculations are described in APPENDIX C. The calculated Fermi surface is shown in Fig.~\ref{fig4}(a), and the sizes are similar to each other for the electron and hole pockets. The superconducting gap is set to $\Delta=1.8 \cos k_x \cos k_y$ meV for $s^\pm$ pairing symmetry, and the gap values on the electron pocket and hole pocket are similar and isotropic, i.e., $-1.668\pm 0.005$ meV for the electron pocket and $1.664\pm0.007$ meV for the hole pocket. To compare with the experimental data, the scalar potential is set to be $V_s=-24$ meV for the non-magnetic impurity, which is relatively weak. Figure~\ref{fig4}(b) shows the calculated two tunneling spectra in the impurity free region and at the non-magnetic impurity site, respectively. The simulated superconducting spectra are very similar to the experimental data with the coherence peak energy at about $\pm 1.5$ meV. The change of coherence peak energy is very little on the spectrum at the impurity site, because $V_s$ is very weak. We calculate the QPI images with a non-magnetic impurity locating at the center of the field of view, see an example in Fig.~\ref{fig5}(a), and then show the resultant FT-QPI patterns in Fig.~\ref{fig4}(c-l). The intensities of characteristic scattering patterns are strongest near $\pm 1.5$ meV due to the strong DOS near the coherence peaks. When the energy is far above the superconducting gap, all the inter-pocket scattering patterns become very weak, which is similar to our experimental data. Then we apply Eq.~\ref{eq2} to the calculated FT-QPI patterns and show the obtained PR-QPI patterns $g_{pr}(\mathbf{q},-E)$ in Fig.~\ref{fig4}(m-q). Clearly, at negative energies near the SC gap, the values of PR-QPI signal $g_{pr}(\mathbf{q},-E)$ are negative around $\mathbf{q}_2$, while they are positive around $\mathbf{q}_3$. This is consistent with the sign change of the gap function on the hole and electron pockets, and it is also consistent with the experimental data shown in Fig.~\ref{fig3}(a)-(e). To obtain the energy evolution of the PR-QPI signal for two different scattering channels $\mathbf{q}_2$ and $\mathbf{q}_3$, we integrate the numerical data near these vectors and show the results in Fig.~\ref{fig4}(r). One can see that the $g_{pr}(\mathbf{q},-E)$ values for the $\mathbf{q}_3$ spots are always positive, while those for the $\mathbf{q}_2$ spots are negative with a peak near $-E=-1.5$ meV. The absolute value of $g_{pr}(\mathbf{q},-E)$ for the $\mathbf{q}_2$ spots decreases with the increase of $E$, and the signal becomes very small and featureless when $E>4$ meV.

All of the theoretical calculation results above are based on the scattering induced by a single non-magnetic impurity with a weak scattering potential in a superconductor with $s^\pm$ pairing, and they agree well with the experimental data. However, the experimental data are not carried out in a region with a single impurity sitting at the center of the image, but in an area with many impurities with weak scattering potential. To compare with the results of a single impurity at the center, we do the calculation by moving the impurity away from the center of the field of view. One example is shown in Fig.~\ref{fig5}(e-h). One can find that the phase-reference pattern [Fig.~\ref{fig5}(h)] calculated at $-1.5$ meV is similar to the one with a impurity at the center as shown in Fig.~\ref{fig5}(d), no matter where the impurity is, which will be addressed later. We also perform calculations for the situation with multiple impurities, and one set of the calculated results are shown in Fig.~\ref{fig5}(i-l). The phase-reference pattern is also similar to the one with a single impurity, and is independent of the number or the positions of the impurities. Hence the DBS-QPI method can be applied to the multi-impurity situation, which is in consistent with the results in Bi2212 \cite{QQGuBi2212}. Finally we also calculate the phase referenced signal from the non-magnetic impurity in a superconductor with $s^{++}$ pairing, and the results at $-E=-1.5$ meV are shown in Fig.~\ref{fig5}(m-p). The scattering potential is the same as that used in the calculation for an $s^\pm$ superconductor. One can find that the PR-QPI signals for the $\mathbf{q}_2$ or $\mathbf{q}_3$ scattering spots are all positive, which is very different from the situation in a sign-reversed $s^\pm$ superconductor.

\section{Discussions}

As presented above, by applying the DBS-QPI method, we have successfully visualized the sign reversal of superconducting order parameter in FeTe$_{0.55}$Se$_{0.45}$, which is well consistent with the $s^\pm$ paring model. The applicability of the DBS-QPI method in this material is based on the impurity bound states within the gap, possibly without clear bound state peaks. This is concluded from the fact that we also see the sign reversal feature of the PR-term $g_{pr}(\mathbf{q},-E)$ at the anergy of $-E=-1.0$ meV. However, we cannot exclude the possibility that some bound state peaks may merge into the coherence peaks, since we have seen a negative peak of integrated $g_{pr}(\mathbf{q},-E)$ at around -1.5 meV. The DBS-QPI method can be carried out near the energies of coherence peaks, which may reconcile the two phase reference methods, namely the DBS-QPI and the HAEM's method. If a non-magnetic point impurity locating exactly at the center of the field of view and the background has a center inversion symmetry, we can prove that the two PR methods are consistent with each other through a simple derivation (see APPENDIX D). However, for HAEM's method, the non-magnetic impurity should locate exactly at the center of the image, the location shift of the impurity by a displacement $\mathbf{r}_0$ will produce a phase shift $\mathbf{q}\cdot \mathbf{r}_0$ for each $\mathbf{q}$. Thus a phase correction is required. The simplest way is just to move the impurity site to the origin of the field of view\cite{DavisFeSe,DuZYNP}. However, such a phase shift is not energy dependent, so it will not affect the phase difference $\theta_{\mathbf{q},-E}-\theta_{\mathbf{q},+E}$. This issue makes the PR-term of DBS-QPI method independent of the location of the impurity, which is also discussed in our previous work \cite{QQGuBi2212}.

In our present study, certainly multi-impurities are involved, perhaps with weak scattering potentials. Through a simple derivation\cite{QQGuBi2212}, we found that the phase correction to a system with multiple impurities is just divide a geometrical dependent summation $\sum_je^{i\mathbf{q}\cdot \mathbf{r}_j}$ which is $\mathbf{q}$ dependent, but energy independent. Therefore, no phase correction is needed for a system with multiple impurities when the DBS-QPI method is applied since the corrected phases for positive and negative energies will cancel each other. However, through numerical simulation\cite{HirschfeldPRB}, it was found that the HAEM's method may not work with diluted density of impurities without phase correction. While it probably works again for dense impurities with a random distribution. In this sense, these two methods have still some difference in treating the data from a system with multiple impurities.

In FeTe$_{1-x}$Se$_{x}$ samples, the ARPES measurements report the superconducting gap of about 2 meV for hole pocket \cite{FeSeTeARPES1,FeSeTeARPES2,FeSeTeARPES3,FeSeTeARPES4}, however, the gap for electron pocket seems to be difficult to measure. One previous report claims that the gap for electron pocket is about 4 meV \cite{FeSeTeARPES3}, which is argued to be consistent with the hump feature on the tunneling spectra in STM measurement \cite{HanaguriQPI}. But the observed hump peak energies are not symmetric about the Fermi level \cite{HanaguriQPI}, and these peaks seem to be absent on the spectra shown in Fig.~\ref{fig1}(b) in our sample. Hence, we argue that the superconducting gap values are similar for both hole and electron pockets. In addition, we have obtained anti-phase PR-QPI signal near the coherence peak. When the energy is near the coherence peak, the differential conductivity is high due to the Bogoliubov dispersion, which may enhance the PR-signal for scattering by the multiple impurities with weak scattering potential. However this should be verified by more solid theoretical and experimental efforts. Here we find that the DBS-QPI method can also work successfully even for the system which does not have bound state peaks, and it shows advantages in at least two aspects. Firstly, the obvious impurity-bound-state peaks are not really necessary, and one does not need to intentionally induce an impurity with proper scattering potential. Secondly, the phase-correction is also not necessary, thus we do not need to know the exact positions of the impurities.

\section{Conclusion}To conclude, we explore the application of a recently proposed DBS-QPI technique to a multi-gap superconductor FeTe$_{0.55}$Se$_{0.45}$. The phase reference-QPI signal shows indeed an anti-phase feature between positive and negative energies for the scattering between hole and electron pockets, namely $\mathbf{q}_2=(0,\pi)$, while that for the scatterings between the electron pockets ($\mathbf{q}_1=(0,2\pi)$ and $\mathbf{q}_3=(\pi,\pi)$) are all in-phase. This gives a direct visualization of the $s^\pm$ pairing model predicted early in the field. Although the DBS-QPI method was originally proposed to treat the data around the bound state peaks, in our present sample, we do not see any obvious bound state peaks, but the method seems also to work well. We argue that this method is applicable to determine the gap structure with a sign reversal in more general cases with even only the multiple non-magnetic impurities of weak scattering potentials.

\section*{APPENDIX A: Experimental methods}

The FeTe$_{1-x}$Se$_{x}$ single crystals with nominal composition of $x = 0.45$ were grown by self-flux method \cite{samplegrowth}. The excess Fe atoms were eliminated by annealing the sample at 400 $^\circ$C for 20 hours in O$_2$ atmosphere followed by quenching in liquid nitrogen. The STM measurements were carried out in a scanning tunneling microscope (USM-1300, Unisoku Co., Ltd.) with the ultra-high vacuum, low temperature and high magnetic field. The samples were cleaved in an ultra-high vacuum with a base pressure about $1\times10^{-10}$ torr. Tungsten tips were used for the STM measurements. A typical lock-in technique was used for the tunneling spectrum measurements with an ac modulation of 0.3 mV and 973.8 Hz. All the experiments in this work are carried at 0.4 K. The FT-QPI images are fourfold symmetrized to enhance the signal-to-noise ratio.

\section*{APPENDIX B: Effect on scattering spot boundary by thickness of Fermi pocket}

\begin{figure}
\includegraphics[width=8cm]{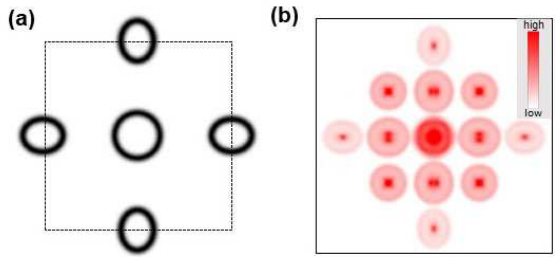}
\caption{(a) Schematic plot of Fermi pockets with the same thickness of the line width as in Fig.~\ref{fig2}(m), but twice larger size of Fermi pockets. (b) Theoretical simulation of FT-QPI pattern by applying self-correlation to (a).} \label{fig6}
\end{figure}

Figure~\ref{fig2}(m,o) shows the schematic images of the Fermi pockets. The sizes of the Fermi pockets are based on the ARPES data \cite{FeSeTeARPES1}, while the the outline-widths of the Fermi pockets are different for different figures. The outline-width of the Fermi pockets is constructed by assuming a Gaussian form of DOS crossing the line. The  Gaussian function can be expressed as $I=I_0\exp(-\delta k^2/2\sigma_k^2)$. Here $I_0$ is the maximum of the DOS intensity, and $\delta k$ is the shortest momentum distance from a certain point to the outline of ideal Fermi pockets. $\sigma_k$ is Gaussian root-mean-square width in momentum space, and $\sigma_k\simeq 0.0056\pi$ for Fig.~\ref{fig2}(m) while $\sigma_k\simeq 0.028\pi$ for Fig.~\ref{fig2}(o). Figure~\ref{fig6}(a) shows the schematic plot of Fermi pockets with the same outline-width as that in Fig.~\ref{fig2}(m), but the sizes of the Fermi pockets are twice larger than those in Fig.~\ref{fig2}(m). Compared with the simulated FT-QPI pattern in Fig.~\ref{fig2}(p), the outer contours of the scattering spots shown in Fig.~\ref{fig6}(b) are much more clear due to the larger sizes of Fermi pockets.

\section*{APPENDIX C: Theoretical calculation details}

We use a simple two-orbital model to calculate the QPI images and FT-QPI patterns. The Hamiltonian in superconducting state \cite{JiangpingHu} is $H=\sum_\mathbf{k}\Psi^\dag (\mathbf{k})B(\mathbf{k})\Psi(\mathbf{k})$ with
\begin{widetext}
\begin{equation}
B(\mathbf{k})=\left(\begin{array}{cccc}
\epsilon_h(\mathbf{k})-\mu_1 & \Delta(\mathbf{k}) & \epsilon_m(\mathbf{k}) & 0 \\
\Delta^{*}(\mathbf{k}) & -\epsilon_h(\mathbf{k})+\mu_1 & 0 & -\epsilon_m(\mathbf{k})\\
\epsilon_m(\mathbf{k}) & 0 & \epsilon_e(\mathbf{k})-\mu_2 & \Delta(\mathbf{k})\\
0 & -\epsilon_m(\mathbf{k}) & \Delta^*(\mathbf{k})& -\epsilon_e(\mathbf{k})+\mu_2
\end{array}\right).
\end{equation}
\end{widetext}
The dispersions of the Fermi surfaces \cite{PRB2011} are as follows. $\epsilon_h=2t_1 (\cos k_x+\cos k_y )-\mu_1$  is for the hole-like pocket near $\Gamma$ point; $\epsilon_e=4 t_2 (\cos k_x \cos k_y )-\mu_2$ is for the electron-like pocket near each M point; $\epsilon_m=t_3 \cos k_x \cos k_y$ is the mixed term of the hole and electron bands. The SC order parameter $\Delta=\Delta_0 \cos k_x \cos k_y$ for the $s^\pm$ pairing, and $\Delta=\left|\Delta_0 \cos k_x \cos k_y\right|$ for the $s^{++}$ pairing. For the numerical work, we take $t_1=40$ meV, $t_2=30$ meV, $t_3=-15$ meV, $\mu_1=153$ meV, $\mu_2=-110$ meV and $\Delta_0=1.8$ meV. The chemical potentials are set to be much larger than the experimental values to reduce the size difference of each Fermi pocket at positive and negative energies. The QPI images are calculated by using the equations and parameters mentioned above, and the FT-QPI and PR-QPI patterns are calculated based on the resultant QPI images.

\section*{APPENDIX D: Comparison of HAEM's method and the DBS-QPI method}

We can consider an ideal situation that a non-magnetic point impurity locates at a geometry symmetrical point on a central symmetric atomic lattice, and the impurity is exactly at the center of a field of view. Then the resultant QPI image should be centrosymmetric, i.e., $g(\mathbf{r},E)=g(-\mathbf{r},E)$. The $g(\mathbf{q},E)$ can be expressed as
\begin{eqnarray}
g(\mathbf{q},E)&=&\sum_{\mathbf{r}}g(\mathbf{r},E)e^{-i\mathbf{q}\cdot\mathbf{r}}\nonumber\\
&=&\frac{1}{2}\sum_{\mathbf{r}}[g(\mathbf{r},E)+g(-\mathbf{r},E)]e^{-i\mathbf{q}\cdot\mathbf{r}}\nonumber\\
&=&\frac{1}{2}\sum_{\mathbf{r}}g(\mathbf{r},E)[e^{-i\mathbf{q}\cdot\mathbf{r}}+e^{-i\mathbf{q}\cdot(-\mathbf{r})}]\nonumber\\
&=&\sum_{\mathbf{r}}g(\mathbf{r},E)\cos(\mathbf{q}\cdot\mathbf{r}).
\end{eqnarray}
Hence, $g(\mathbf{q},E)$ will always be real, namely, the phase of $g(\mathbf{q},E)$ is either 0 or $\pi$. For the HAEM's method, the PR-term in Eq.~\ref{eqHAEM2} can be then rewritten as
\begin{eqnarray}
 &\delta \rho^-(\mathbf{q},E) = g(\mathbf{q},+E)-g(\mathbf{q},-E)\nonumber\\
&= \mathrm{Re}[g(\mathbf{q},+E)-g(\mathbf{q},-E)]\nonumber\\
&= |g(\mathbf{q},+E)|\cos\theta_{\mathbf{q},+E}-|g(\mathbf{q},-E)|\cos\theta_{\mathbf{q},-E}.
\end{eqnarray}
For the sign-reversal gaps, $\delta \rho^-(\mathbf{q},E)$ will be manifestly enhanced at the energies between $\Delta_1$ and $\Delta_2$ according to HAEM's method. Since the phase of $g(\mathbf{q},\pm E)$ can only be 0 or $\pi$, one can expect that $|\theta_{\mathbf{q},-E}-\theta_{\mathbf{q},+E}|=\pi$, and $\delta \rho^-(\mathbf{q},E) \sim  |g(\mathbf{q},+E)|+|g(\mathbf{q},-E)|$ is a large positive value. Then, according to eq.\ref{eq2}, $g_{pr}(\mathbf{q},-E) = -|g(\mathbf{q},-E)|$, which is exactly the result given by DBS-QPI method for sign-reversal gaps. For the sign-preserved gaps, the HAEM's method predicts that $\delta \rho^-(\mathbf{q},E)$ will be very small at the energies between $\Delta_1$ and $\Delta_2$. In the ideal case of the model, that means $\theta_{\mathbf{q},-E}=\theta_{\mathbf{q},+E}$, and $\delta \rho^-(\mathbf{q},E) \sim |g(\mathbf{q},+E)|-|g(\mathbf{q},-E)|$ is a negligible value. According to eq.\ref{eq2}, $g_{pr}(\mathbf{q},-E) = |g(\mathbf{q},-E)|$ is positive, which is also consistent with the conclusion of DBS-QPI method for sign-preserved SC gaps. In this point of view, we argue that the two PR-QPI methods consist with each other about the gap sign problem in a qualitative way. This further indicates that the phase difference of the FT-QPI signal at positive and negative energies observed by experiments comes really from the sign change of the superconducting gaps, but not due to the particular method applied here.

\begin{acknowledgments}
We acknowledge useful discussions with Qimiao Si. The work was supported by National Key R\&D Program of China (grant number: 2016YFA0300401), National Natural Science Foundation of China (NSFC) with the project: 11534005.
\end{acknowledgments}

Mingyang Chen and Qingkun Tang contributed equally to this work.

$^*$ huanyang@nju.edu.cn

$^\dag$ qhwang@nju.edu.cn

$^\ddag$ hhwen@nju.edu.cn

\end{document}